\documentclass{aa}

\usepackage{graphics}

\newcommand{\Msun}{$M_{\odot}$}
\newcommand{\he}{He$\,$2-436}
\newcommand{\wray}{Wray$\,$16-423}
\newcommand{\Teff}{\mbox{$T_{\rm eff} $}}
\def \hii{\ion{H}{ii}}
\def \hi{\ion{H}{i}}
\def \hei{\ion{He}{i}}
\def \heii{\ion{He}{ii}}
\def \cii{\ion{C}{ii}}
\def \ciii{\ion{C}{iii}]}
\def \civ{\ion{C}{iv}}

\def \oi{[\ion{O}{i}]}
\def \oii{[\ion{O}{ii}]}
\def \oiii{[\ion{O}{iii}]}
\def \sii{[\ion{S}{ii}]}
\def \ariii{[\ion{Ar}{iii}]}
\def \ariv{[\ion{Ar}{iv}]}
\def \eg{{\it e.g.}}
\def \etc{{\it etc}}

\begin{document}

   \thesaurus{01             
             (02.14.1        
              09.16.1        
              09.16.1 He~2-436     
              09.16.1 Wray~16-423  
              11.04.2        
              11.09.1 Sagittarius) 
}

\headnote{Letter to the Editor} 

   \title{Third-dredge-up oxygen in planetary nebulae 
\thanks{Based on observations collected at ESO La Silla, Chile.}
}


  \author{      D.~P\'equignot \inst{1}
          \and
                J.~R.~Walsh \inst{2}
          \and
                A.~A.~Zijlstra \inst{3}
          \and
                G.~Dudziak \inst{4}
          }

   \offprints{D.~P\'equignot}

   \institute{  Laboratoire d'Astrophysique Extragalactique et de 
                Cosmologie associ\'e au CNRS (UMR 8631) et \`a l'Universit\'e 
                Paris 7, DAEC, Observatoire de Paris-Meudon, F-92195 
                Meudon C\'edex, France. Email: daniel.pequignot@obspm.fr 
         \and
                Space Telescope European Co-ordinating Facility, 
                European Southern Observatory, Karl-Schwarzschild Strasse 2, 
                D-85748 Garching bei M\"unchen, Germany. E-mail: jwalsh@eso.org
         \and
                Department of Physics,
                University of Manchester Institute of Science and Technology, 
                P.O. Box 88, Manchester, M60 1QD, United Kingdom. 
                Email: aaz@iapetus.phy.umist.ac.uk 
         \and
                Department of Physics and Applied Physics, University of 
                Strathclyde, Scotland. E-mail: gregory.dudziak@strath.ac.uk. 
             }

   \date{Received ? / Accepted ?}

   \maketitle

   \begin{abstract}
The planetary nebulae He$\,$2-436 and Wray$\,$16-423 in the Sagittarius 
dwarf galaxy appear to result from nearly twin stars, except that 
third-dredge-up carbon is more abundant in He$\,$2-436. A thorough 
photoionization-model 
analysis implies that ratios Ne/O, S/O and Ar/O are significantly 
smaller in He$\,$2-436, indicative of third-dredge-up oxygen enrichment. 
The enrichment of oxygen with respect to carbon is (7$\pm$4)\%.  
Excess nitrogen in Wray 16-423 suggests third dredge-up of late-CN-cycle 
products even in these low-mass, intermediate-metallicity stars. 

   \end{abstract}


\section{Introduction}
\label{sec:intro}

It is usually assumed (\eg\ Henry \cite{henry89}) that 
the oxygen abundance of a planetary nebula (PN) reflects the metallicity $Z$ 
of the interstellar medium (ISM) in which the parent star formed, inasmuch 
as the conversion of $^{16}$O into $^{14}$N due to advanced CNO 
processing is unimportant, a condition fulfilled in low-mass stars 
($M <$~2~\Msun). Low-mass stars bring to their surface nitrogen 
(through first dredge-up), produced by CN-cycle conversion of carbon, 
and carbon (third dredge-up), produced by 3-$\alpha$ fusion along the 
Asymptotic Giant Branch (AGB, \eg\ Iben \& Renzini \cite{iberen83}). 
During He-burning, $^{14}$N is transformed notably into $^{22}$Ne 
and some $^{12}$C into $^{16}$O. Thus third dredge-up is likely to 
bring freshly synthesized oxygen to the surface of low-mass stars, as 
suggested by Boothroyd \& Sackmann (\cite{boosac88}, here BS88) and, 
in a preliminary form, by Herwig et al. (\cite{herblo00}, here HBD), 
who invoked intershell dredge-up (``fourth dredge-up'', Iben \cite{iben99})  
by diffusive overshoot. Exposed deep layers of hydrogen-deficient post-AGB 
stars appear to have quite similar concentrations of oxygen and carbon 
(Koesterke \& Hamann \cite{koeham97}; Herwig et al. \cite{herblo99}) 
but {\sl the amount of fresh oxygen expelled from AGB stars is unknown}. 
The oxygen abundance is difficult to determine in the atmosphere of 
evolved stars (\eg\ Fulbright \& Kraft \cite{fulkra99}) and shows 
large scatter in PNe of a given galaxy (\eg\ Leisy \& Dennefeld 
\cite{leiden96}, here LD96). 

In order to reveal the oxygen enrichment of PNe, one must seek for 
circumstances enabling the detection of a small excess of oxygen. 
Two PNe, \he\ and \wray, were found (Zijlstra \& Walsh \cite{zijwal96}, 
here ZW96; Walsh et al. \cite{waldud97}) to belong to the Sagittarius dwarf 
galaxy (Ibata et al. \cite{ibagil95}). Spatial and kinematic properties 
point to a common origin for these PNe. Photoionization models 
(Dudziak et al. \cite{dudpeq00}, here DPZW) indicate nearly 
identical depletions with respect to solar for all elements beyond 
nitrogen [($-0.55\pm0.07$)~dex]. Both PN nuclei are early-type [WC] 
stars belonging to nearly the same (H-burning) evolutionary track 
$M$=($1.2\pm0.1$)\Msun, $Z$=0.004 (Vassiliadis \& Wood \cite{vaswoo94}),
consistent with their precursors being the intermediate-age carbon stars 
of Sagittarius. As expected, third dredge-up yield was large 
in these intermediate-$Z$ stars (Marigo et al. \cite{margir99}), 
but even larger in \he. If both stars were born in the same 
star formation episode with identical abundances, advantage 
can be taken of their low initial abundances and different 
third-dredge-up yields to study the influence of this process upon 
oxygen and other elements by comparison of the respective abundances. 
Here, unlike the general study of DPZW, emphasis is on abundance 
{\sl ratios} and differential effects. 
P\'equignot et al. (\cite{peqzij00}) gave a preliminary account. 

\section{The Sagittarius dwarf galaxy PN models} 
\label{sec:pne}

Spectra of  \wray\ and \he\ were secured in identical conditions. 
Their small apparent size led to global spectra 
of the highest reproducibility and relevance for modeling. 
The many theoretical line ratios (\hi, \hei, forbidden multiplets) 
allowed to check that statistical errors on measured fluxes were 
realistic, the few discrepancies being indeed similar in both PNe. 
Thus, in a direct comparison, systematic errors are likely to cancel out. 

In the photoionization model analysis of DPZW, the error bars 
attached to the abundances of these PNe were estimated from comparing 
two independent codes: they represented, at some "1$\sigma$ level", 
the uncertainties left after evaluating the origin of the 
differences between codes and trying to include systematic errors. 
Here, only one of the photoionization codes 
(code NEBU; \eg\ Morisset \& P\'equignot \cite{morpeq96}) is used, 
considering only the statistical error on line fluxes. Because of the 
neglect of systematic errors, the intervals are much reduced. Conversely, 
because of the "plasticity" allowed by the rather large number of 
free parameters and the thorough exploration of all possible solutions, 
the intervals can be enlarged. The abundance {\it ratios}, 
which were not considered in the error analysis of DPZW,
are determined with greater accuracy than the individual abundances,
as shown below. 

Current [WC]-atmosphere models are not reliable: 
in order not to miss solutions because of restrictive assumptions, 
black-body stars with a discontinuity at the He$^+$ limit were 
employed as a three-parameter sequence and an equivalence was established 
with the NLTE model stellar atmosphere (effective temperature \Teff), 
from the grid of Clegg \& Middlemass (\cite{clemid87}). 
Similar free-parameter sets, procedures and convergence criteria 
were adopted to obtain two-sector models for both PNe. 
The models matched essentially all line fluxes within 1$\sigma$ errors, 
that is no more than a few percent of line strength for most basic lines. 
The two discrepancies left, 
one of them involving \oi\ and the other one \ariv/\ariii, 
turned out to be identical for both PNe, suggesting systematic effects. 

It was possible to generate several models of each PN satisfying 
the very strong and numerous observational constraints, as three 
key line fluxes (\cii~426.7, \oii~372.6 and \sii~406.8$nm$) were 
not very accurately determined. Given a \Teff\ for the central star, 
the model parameters are constrained by imaging and/or radio continuum data 
and by (accurate) optical line fluxes. A noticeable aspect is the global 
energy balance. The cooling rate, mainly determined by the 
unknown \ciii\ and \civ\ UV line fluxes (in addition to \oiii), 
depends on the carbon abundance, to which \cii\ is proportional. 
For \wray, the computed \cii\ flux is always in agreement with observation, 
stronger limitations being brought by other lines. For \he, considering 
different models with increasing density in the dense sector 
(an incomplete inner shell), both the \oiii500.7$nm$/\oiii436.3$nm$ ratio 
and the \oiii500.7$nm$ flux tend to decrease, but can be reconciled 
with observation by increasing the oxygen and carbon abundances. 
As free parameters ensure sufficient flexibility, 
the uniqueness of the solution breaks down 
and the boundaries of the set of acceptable models (and abundances) are 
now determined by the \cii\ flux. 

In Fig.~1 are shown for both PNe the ranges of allowed carbon and oxygen 
abundances versus the \Teff's. Varying \Teff\ makes the predicted 
\sii~406.8$nm$ and \oii\ 372.6$nm$ fluxes vary relative to other lines 
in models. For \he, low or high \Teff's are forbidden as the 
\sii\ or \oii\ fluxes become too large. Relaxing somewhat the error bars 
attached to these lines increases the allowed range of \Teff\ without 
much changing the range of acceptable abundances. As for \wray, 
all elements but carbon have abundances virtually independent of \Teff. 
Low \Teff's (low C abundances) are forbidden by two independent \sii\ lines 
being predicted too weak; high \Teff's by one accurately observed \sii\ line 
becoming too large. Higher \Teff's are also excluded by the fact 
the He$^+$ discontinuity, constrained by \heii~468.6$nm$, would be 
much larger than in atmosphere models. 
Also the shape of the model nebula would then deviate from observation. 

The oxygen abundances of the \he\ models encompass the well 
defined value of \wray. Nevertheless oxygen {\sl tends} to be larger in 
\he. As a check of this tendency, \wray\ models were run with the oxygen 
abundance increased by 10\%. No satisfactory solution could be found. 
Re-convergence on all parameters was performed until all lines 
except \oiii~500.7$nm$ and \oiii~436.3$nm$ were again satisfactory and 
the discrepancy on both \oiii\ line fluxes was reduced to a minimum, 
typically 5$\sigma$. Other elemental abundances increased by 
9$\pm$1\% (except C: 18\%). An \he\ model of given \Teff\ was then 
re-converged to a solution, though with the condition that the computed 
\oiii\ line fluxes should reproduce the discrepancy met in the previous 
\wray\ model. In this new \he\ model, most abundances were shifted upwards 
relative to the standard solution by 8$\pm$1\% (except N: 3\% and C: 18\%), 
close to the shift imposed on oxygen in the \wray\ model. 
This numerical experiment suggests that systematic errors 
may substantially shift the derived abundances but not influence 
conclusions based on differences between the two PN models. 

\begin{figure}[ht]
\resizebox{\hsize}{!}{\includegraphics{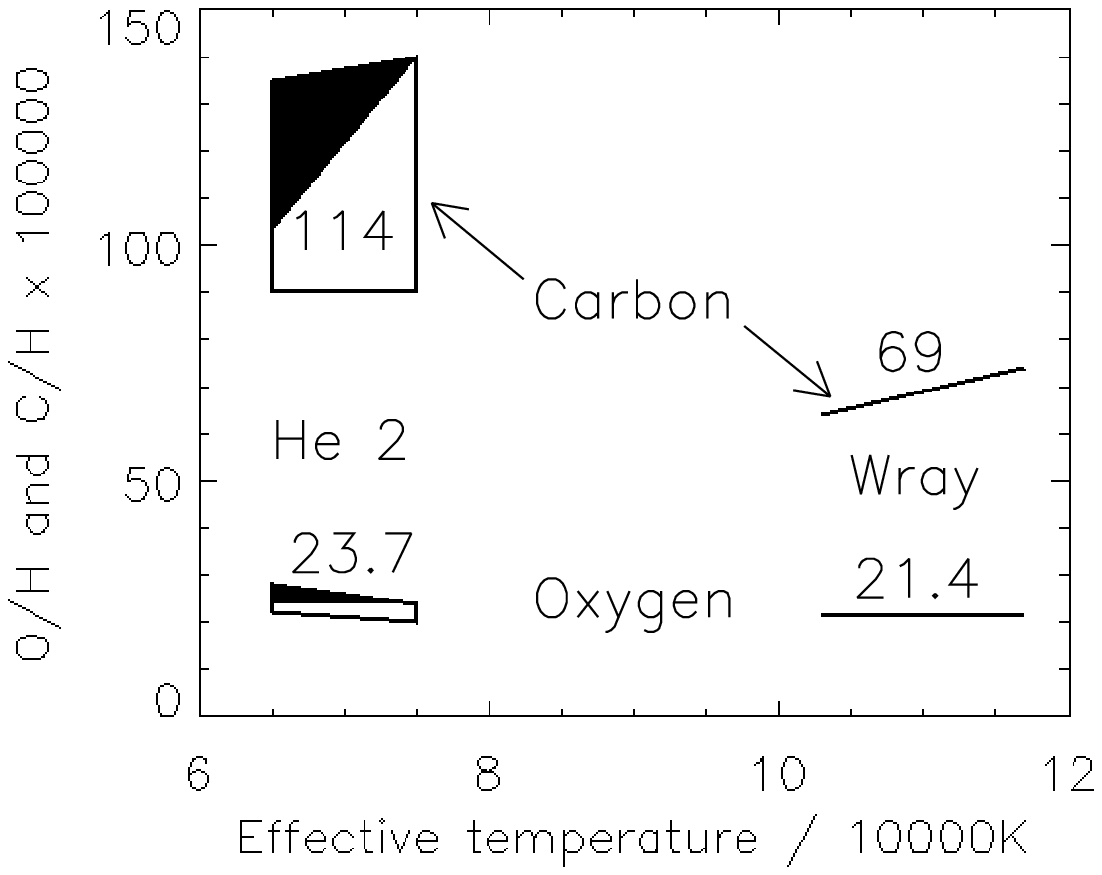}}
\caption[]{Acceptable abundances of C and O by number versus \Teff\ of central stars according to model nebulae for \he\ and \wray. Best-model values are indicated. The shaded areas correspond to \he\ models having the same sulfur abundance as \wray\ within uncertainties (Sect.~3). For \Teff/10$^4$K = 6.5, 10$^5$$\times$O/H increases from 24 to 28 as 10$^5$$\times$C/H increases from 104 to 135. For \Teff/10$^4$K = 7.5, the O and C values are 24 and 140 respectively.}
\end{figure}

Table~1 lists the absolute helium and oxygen abundances by number 
and other abundances relative to oxygen. 
For all elements but carbon, the quoted abundance ``uncertainty'' 
is obtained by adding linearly the full abundance spread of acceptable 
models to the 1$\sigma$ error of the best determined line 
(the models were constrained to reproduce this line exactly). Percent 
differences on abundance ratios and  1$\sigma$ error bars (Table~1, col.~4) 
clearly show that Ne/O, S/O and Ar/O are smaller in \he\ than in \wray\ 
(3.5, 1.8 and 1.4$\sigma$ confidence levels respectively, hence 
4.2$\sigma$ collectively). Given that sulfur and argon are not synthesized 
in low-mass stars and that neon should hardly be affected either, 
{\sl this very significant trend 
must be due to a larger oxygen abundance in \he}. An explanation involving 
an abundance fluctuation in the ISM with, \eg, an excess {\sl specific} 
to oxygen at the birth place of the parent star of \he\ is unlikely, 
as these stars are twins in most respects (ZW96, DPZW) {\sl and} 
the Ne/O ratio is remarkably stable, at $\sim$~0.18, in the ISM of even 
different galaxies (\eg\ Henry \cite{henry89}). Thus the most plausible 
interpretation of the excess is in terms of {\sl oxygen production}, with the 
precursor star of \he\ having been more efficient than that of \wray. Since 
\he\ shows a larger carbon excess --~the signature of third dredge-up~-- 
it is inferred that this process was also able 
to bring freshly synthesized oxygen to the star surface. 

\begin{table}
  \caption{Abundance ratios by number in Sagittarius PNe}
\begin{flushleft}
    \begin{tabular}{lccc} \hline
Ratio &     \wray\        &        \he\    & diff. (\%)\\ \hline
He/H &\ 0.108$\pm$0.002 \ & 0.108$\pm$0.003 & \ \ 0.0$\pm$3.6 \\
O/H  &(21.4$\pm$0.2)$\times$10$^{-5}$&(24.0$\pm$4.0)$\times$10$^{-5}$ 
                                            & --11.5$\pm$16. \\ \hline
C/O  & 3.22$\pm$0.23       & 5.0$\pm$0.8    & --43.3$\pm$17. \\
N/O  & 0.213$\pm$0.015   & 0.116$\pm$0.006  & 59.0$\pm$8.7 \\
Ne/O & 0.168$\pm$0.003   & 0.147$\pm$0.005  & 13.3$\pm$3.8 \\
S/O  & 0.0201$\pm$0.0008 & 0.0162$\pm$0.0016& 24.0$\pm$13. \\
Ar/O & 0.0042$\pm$0.0004 & 0.0029$\pm$0.0007& 36.6$\pm$26. \\ \hline
    \end{tabular}
\end{flushleft}
\end{table}

\begin{table}
  \caption{Evolution of abundances$^\ast$ in stellar envelopes}
\begin{flushleft}
    \begin{tabular}{lcccc} \hline
Stage:&Init&1st~drge-up &3rd~drge-up &3rd~drge-up \\
Elmt&  & + CBP$^a$ &(Wray) & (He 2) \\ \hline
He& 8250& 9000 &10800$\pm$200 \ \ &10800$\pm$300 \ \  \\
O &(17)&16.5$\pm$3.2 \ &21.4$\pm$0.2 \ &26.0$\pm$2.0 \ \\ 
C & 4&2.5&69.$\pm$5. \  &122.$\pm$18. \ \\
N & 1&2.5&4.6$\pm$0.3&3.0$\pm$0.3 \\
Ne& -& 3.4 &3.59$\pm$0.07&3.8$\pm$0.3 \\
S & -& 0.43 &0.43$\pm$0.02&0.43$\pm$0.02 \\
Ar& -& 0.08 &0.09$\pm$0.01&0.07$\pm$0.02 \\ \hline
    \end{tabular}

$^\ast$Abundances by number in units H = 10$^5$\\
$^a$Cool Bottom Processing (Boothroyd \& Sackmann \cite{boosac99})\\
\end{flushleft}
\end{table}

\section{Discussion} 
\label{sec:disc}

In Table~2 is sketched 
a possible evolution of the stellar envelope composition, assuming that 
(1) initial compositions were identical in the \wray\ and \he\ stars, 
(2) sulfur abundance was constant, keeping the value 
determined for \wray\ within uncertainties (col.~4 of Table~2), and 
(3) oxygen abundance varied linearly with that of carbon  
during third dredge-up (cols.~4, 5). 
Sulfur is preferred to argon as a reference constant-abundance element, 
in view of the moderate scatter of the sulfur abundance 
determinations in models and the difficulty met in accounting for 
the argon ionization (DPZW). Scatter is less for neon, but 
its nucleogenic status may be less clear-cut and the variation 
of Ne/O from \wray\ to \he, although significant, is smaller. 
Reliable sulfur abundances are difficult to obtain in PNe, yet 
this approach is justified considering that models are based on 
homogeneous procedures and used only in a differential manner. 

Once the sulfur abundance is specified for \he, the range of possible 
oxygen abundances is narrowed: assuming the errors are independent, 
S/O (Table~1, col.~3) and S (Table~2, col.~4) imply 
23.6 $<$ 10$^5$$\times$(O/H) $<$ 28, an interval that excludes the 
O/H value of \wray. Owing to the linear increase of S with O in 
the model results, the minimum O/H is in fact slightly larger 
(Table~2, col.~5; Fig.~1, shaded areas). 
Let $\Delta$O and $\Delta$C be the increments by number 
of the oxygen and carbon abundances due to third dredge-up. 
Taking into account correlations between C and O in \he\ models 
(Fig.~1: the lowest C goes with the lowest O) and assumption (3) above, 
the difference in abundances between the two PNe shows that extreme 
values for $\Delta$O/$\Delta$C are 2.4/76=0.03 and 6.8/61=0.11. 

The extrapolated main-sequence abundance of oxygen (Table~2, col.~3 or 2), 
obtained assuming carbon is initially much less abundant than in the PNe, 
is used only to determine $Z$ and relevant initial abundances (see below). 

The present result for $\Delta$O/$\Delta$C seems to fit with 
old predictions of BS88 for flash-produced intershell O/C in the 
region mixed after convection but this may be fortuitous: these 
stellar models are uncertain, as indicated by Herwig (\cite{herwig00}) 
who obtained models for intermediate-mass high-$Z$ stars 
using an exponential diffusive overshoot formalism. 
Based on the same concept, HBD obtained preliminary 
models for low-$Z$ stars, suggesting that surface O/C reaches 
asymptotically the intershell value, $\sim$~1/3. 
Thus, considering tendencies shown by recent star models, 
the present coarse estimate would seem to imply that 
{\sl relatively little oxygen} should be third-dredged-up. 
Nonetheless a PN is formed after many ejection episodes and 
its abundance may be far from reflecting asymptotic values. 

By contrast, neon and argon appear constant within uncertainties 
and the lower abundance of nitrogen in \he\ cannot be due to 
extensive conversion of this element into heavier ones. 
The initial (Table~2, col.~2) and first-dredge-up (col.~3) 
helium, carbon and nitrogen abundances are from 
Boothroyd \& Sackmann (\cite{boosac99}) assuming $M$=1.2~\Msun\ 
and $Z$=0.004 (in keeping with extrapolated oxygen). Second dredge-up 
does not occur in low-mass stars, but these authors 
take into account deep-circulation mixing leading to 
enhanced $^{13}$C/$^{12}$C, as observed along the red giant branch, 
and to further conversion of initial carbon into nitrogen. This so-called 
``Cool Bottom Processing'', sensitive to $Z$ and $M$, can bring 
the nitrogen abundance quite close to that of \he\ but 
perhaps not to the larger one of \wray\ (Table~2). 
Another source of nitrogen is required. 

Mixing of $^1$H with fresh $^{12}$C occurs in low-mass AGB stars 
to produce $^{13}$C, s-nuclei \etc, but the hydrodynamics involved is 
not understood (Lattanzio \& Frost \cite{latfro97}). It may not be 
excluded that some $^{13}$C can find its way up to the hydrogen shell 
or that insertion of H-rich pockets into the 
He/C-rich intershell is, on occasion, efficient enough to allow 
CN-cycle completion. In this way, low-mass AGB stars could 
synthesize some $^{14}$N from fresh $^{12}$C and mimic 
intermediate-mass stars in which the so-called ``Hot Bottom Burning'', 
together with third dredge-up, is believed to lead to (N-rich) Type~I PNe. 
Within the uncertainties of the definition 
(Torres-Peimbert \& Peimbert \cite{torpei97}), \wray\ would in fact 
be marginally Type~I in the Large Magellanic Cloud (LMC). The indication 
here is that {\sl the dredge-up of additional nitrogen was more effective 
in \wray, where carbon is less overabundant}. 

According to BS88, the He/C ratio in the carbon pockets that will be 
dredged-up levels off rapidly at He/C $\sim$ 8 by number, compatible 
with the ``observed'' third-dredge-up $\Delta$He/$\Delta$C, which 
lies anywhere between 15 and $-15$ (Table~2, cols.~4, 5), but not 
with the large jump of helium between first dredge-up and \wray. 
This may suggest, among other possibilities, that the initial helium 
abundance adopted here for these stars was too small. 

Abundances are comparable in the PNe of Sagittarius and LMC (LD96). 
The abundance scatter in the LMC is likely due to the coexistence of 
several populations, but Ne/O is expected to be relatively stable, 
except possibly for differential effects of the kind revealed here. 
>From data of LD96, it is found that log(Ne/O) is 
--0.61$\pm$0.10 and --0.83$\pm$0.10 for Type-I and non-Type-I PNe 
respectively. These determinations bracket the mean \hii-region value, 
--0.74 (see LD96), and appear marginally different. Assuming that neon 
best traces the overall LMC evolution, this suggests that oxygen is 
relatively depleted in Type~I, at least as a consequence of 
second dredge-up in intermediate-mass stars (see also  
Dopita \& Meatheringham, \cite{dopmea91}), but {\sl enhanced} in non-Type~I 
by third dredge-up in accord with present finding in Sagittarius PNe. 

Abundances are lower in the SMC than in the LMC. Nitrogen-rich PNe 
of the SMC (type $i$+I of LD96) are found among extremely 
low $Z$ stars, presumably of low initial mass. Also some halo PNe 
have N/O $>$ 1 (Howard et al. \cite{howhen97}), at variance with the 
well-known association of Type~I PNe with intermediate-mass stars, 
but maybe another manifestation of what is occuring in Sagittarius: 
{\sl $^{14}$N is enhanced by third dredge-up in low-mass AGB stars} 
and, due to a combination of efficiency and titration effect, 
N/O can reach large values for sufficiently small $Z$. 

The discussion of the Sagittarius PN models (DPZW) makes it clear that 
improvement of the critical carbon abundance can be expected from more 
complete observations (UV, far red). Nevertheless the present results 
can already provide some constraints on the new generation of models for 
the evolution of low-mass low-$Z$ stars and lead to a re-interpretation 
of trends shown by PN abundances in low-$Z$ environments. The production 
of oxygen by low-mass stars ($\sim$~2$\times$10$^{-4}$\Msun/star in 
Sagittarius) may have to be considered in galactic-evolution models.

\end{document}